\documentclass[letterpaper,twocolumn,superscriptaddress,floatfix]{revtex4}
\usepackage[latin1]{inputenc}
\usepackage{bm}
\usepackage{multirow,amssymb,amsbsy,amsmath}
\usepackage{graphicx}
\usepackage{verbatim}
\makeatletter
\usepackage{pifont}
\makeatother

\begin{document}

\title{Preserving Entanglement of Flying Qubits in Optical Fibers by \\Dynamical Decoupling}
\author{Bin Yan}
\affiliation{Key Laboratory of Quantum Information, University of
Science and Technology of China, CAS, Hefei, 230026, People's
Republic of China}
\author{Chuan-Feng Li$\footnote{email:cfli@ustc.edu.cn}$}
\affiliation{Key Laboratory of Quantum Information, University of
Science and Technology of China, CAS, Hefei, 230026, People's
Republic of China}
\author{Guang-Can Guo}
\affiliation{Key Laboratory of Quantum Information, University of
Science and Technology of China, CAS, Hefei, 230026, People's
Republic of China}

\date{\today }
\begin{abstract}
We theoretically investigate the influence of dynamical decoupling
sequence in preserving entanglement of polarized photons in
polarization-maintaining birefringent fibers(PMF) under a classic Gauss 1/f noise. We study the dynamic evolution of entanglement along the control sequence embedded fibers. Decoherence due to dispersion of polarization mode in PMF can be dramatically depressed, even for a wild optical width. Entanglement degree can be effectively preserved while the control pulse is implemented.
\end{abstract}
\maketitle

Entanglement plays a central role in quantum communication process. Due to an interaction with the uncontrollable degree of freedom of the environment, quantum system may lose its entanglement degree. In an optical fiber-based quantum communication channel, the residual optical birefringence randomly accumulating along the fiber set up a main obstacle to maintain the fidelity of information. Polarization-entangled photons pairs distributed over optical fibers even suffer the process of abrupt disappearance of entanglement\cite{1}, known as a phenomenon of entanglement sudden death\cite{2,3}. Thus it is upper most important to find effective methods for preserving entanglement of polarized photons propagating in optical fibers.

Several approaches have been developed to address this issue. Notable examples are decoherence-free sbuspace\cite{4,5} and quantum error-correction codes\cite{6,7}, both based on carefully encoding the quantum information into a wider, while partially redundant, Hilbert space. Alternative approach is quantum feedback\cite{8}. In this technology information channel is designed to be closed-loop with appropriate measurements and real-time correction to the system. However, all these  strategies have the drawback of requiring a large amount of extra resources. On the contrary, an open-lope control method known as dynamical decoupling avoids all these hindrances.

Dynamical Decoupling(DD) is a simple and effective method for
coherence control. In this technology undesired effects of the
environment are eliminated via strong and rapid time-dependent
pulses faster than the environment correction time. The physical
idea behind DD scheme comes from refocusing techniques in Nuclear
Magnetic Resonance (NMR) systems\cite{9} and then be extended to any
other physical contexts in the last decades, such as
nuclear-quadrupole qubits and electron spins qubits\cite{10}. Preserving of entanglement between two stationary qubits by dynamical decoupling also has been wildly discussed\cite{11,12,13}. Some
prominent examples of DD schemes are the wildly used Hahn's spin
echo, periodic DD (PDD), Carr-Purcell-Meiboom-Gill (CPMG)\cite{14},
concatenated DD (CDD)\cite{15}, Uhrig DD (UDD)\cite{16}, and
recently proposed near-optimal decoupling (QDD)\cite{17} used to
eliminate general decoherence of qubits.

Extension of the time-dependent DD sequences to the spare-dependent dynamic evaluation process lead to the idea of applying DD controls into the optical fiber-based quantum communication channel\cite{18,19}. Experimental implementation has notably revealed the potential of this extension\cite{20,21}. However, this delicately designed experiment is limited to the condition that the noises are systematically introduced in a non-stochastic way, while in a real fiber-based channel they are distributed randomly. In this paper, we investigate the performance of CPMG sequence in preserving entanglement of polarized photons under stochastic classic Gauss noises. We consider here the entangled photons are distributed through polarization-maintaining(PM) birefringent fibers, in which case the polarized photons suffer a pure dephasing process, $ T1>>T2 $. Such building blocks of coherence control for a dephasing dynamic can be extended to general decoherence processes in ordinary single mode optical fibers with more complicated DD schemes like CDD\cite{15} or QDD\cite{17}. In the following, we show the derivation of entanglement evolution using the filter-design method\cite{22} and give the numerical simulation with some special initial states.

We consider the situation of photon distribution depicted in Fig. 1. Photon A is preserved by a quantum register, while the entangled photon B, carrying the encoded information propagates through the fiber to the recipient.
The fiber is embedded with $ \pi $ pules realized by half-wave plates. In our scheme the intervals between waveplates are fixed to a certain scale in the fiber and the waveplates sequence are arranged as continues CPMG cycles. Note that the CPMG sequence has a self-construct feature as illustrated in Fig.1. In other words we can view two cycles of CPMG sequences with N pules each as one cycle of CPMG sequence with 2N pules. This feature of CPMG sequence allows us to analysis the dynamic evaluation of entanglement through the fiber directly, as illustrated in the following.
\begin{figure}[h]
\includegraphics[scale=0.16]{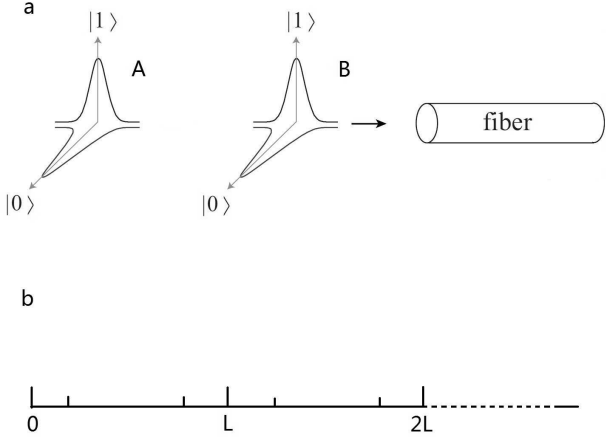}
\caption{(a)Photon distribution scheme. (b)The length between pulses embedded in the fiber is fixed to a certain scale. Measuremnet at length L in the fiber corresponds to a 2 pulses CPMG sequence, while measurement at length corresponds to a 4 pulses CPMG sequence.}
\end{figure}

Under a photon distribution process described above, the qubits-environment interaction Hamiltonian can be written as\cite{23}:
\begin{equation}
\hat{H}=\dfrac{1}{2}\int d\omega(|\omega_{A}><\omega|\otimes b(\omega,L)\sigma_z^A+|\omega_{A}><\omega|\otimes I_B)
\end{equation}
In nondispersive media only the first order of $ \omega $ remains in $ b(\omega,L) $\cite{23}. Thus we can rewrite b as:
\begin{equation}
b(\omega,L)=\omega(\Omega+\beta(L))
\end{equation}
$ \Omega $ is a constant and $ \beta(L) $ presents for the stochastic fluctuation of the noises with a zero mean, and it has a two-point correlation function:
\begin{equation}
S(L_1-L_2)=<\beta(L_1),\beta(L_2)>
\end{equation}
Where $ <> $ corresponds to the average with respect to the noise realizations. The statistical properties of the environment can also be expressed as the spectral density of noise:
\begin{equation}
S(\omega)=\int e^{iwt}S(L)dL
\end{equation}
In the following discussing the statistics of fluctuations is assumed to be Gaussian, in which case the noise is completely defined by the first-order correlation function S(L) in equation(3).
We assume the polarization entangled photon pairs are generated via spontaneous parametric down conversion. The quantum state of the generated photon pairs can then be written as\cite{1}:
\begin{equation}
|\varphi_{in}>=\int d\omega f(\omega)|\omega,-\omega>\otimes|P>
\end{equation}
$ |\omega,-\omega> $ is the frequency basis vector of photon pairs, $ \omega $ denotes the offset from the central frequency. $ |P> $ represents polarization modes, which can be expressed as the superimposing of horizontal and vertical polarization basises $ |HH>,|HV>,|VH>,|VV> $

Due to the stochastic dephasing noise described in Hamiltonian (1), the output state of the photon pairs after photon A propagating through the fiber for a length L without DD sequence implemented is:
\begin{eqnarray}
\nonumber |\varphi_{out}>=\int d\omega f(\omega)e^{-\frac{1}{2}\int_0^L b_s(\omega,L)dL}|\omega,-\omega>\\ \nonumber \otimes (a|HH>+b|HV>)\\ \nonumber
+\int d\omega f(\omega)e^{\frac{1}{2}\int_0^L b_s(\omega,L)dL}|\omega,-\omega>\\\otimes (c|VH>+d|VV>)
\end{eqnarray}
where the subscript s stands for a given realization of noise before taking the ensemble average.
The density matrix that characterizes the detected polarization qubits can be abtended by a process of tracing over the frequency modes and taking ensemble average of the environment noise:
\begin{equation}
\rho=\int D\beta(Tr^\omega|\psi_{out}><\psi_{out}|))
\end{equation}
The integration is a Gaussian functional integral with the variable $ \beta(L) $ having a Gaussian distribution.

The elements of the resulting density matrix are then given by:
\begin{eqnarray}
\nonumber &&\rho_{ii,12,34}=\rho_{ii,12,34}(0),\\
\nonumber  &&i=1,2,3,4 \\
\nonumber&&\rho_{ij}=\rho_{ij}(0)\dfrac{1}{\surd{1+\sigma^2f(L)}}exp({-\dfrac{\omega_0^2f(L)}{1+\sigma^2f(L)}} ),\\
&& i,j=others,
\end{eqnarray}
where $ f(L) $ can be expressed as the integral of the noise spectrum $ S(w) $:
\begin{equation}
f(L)=\int\dfrac{d\omega}{2\pi}S(\omega)\dfrac{F(\omega L)}{\omega^2}
\end{equation}
$ F(x) $is a filter function depends on the shape of DD sequence. For a free evaluation without control implemented $ F(x)=2\sin^2{\dfrac{x}{2}} $.
The fiber-based DD control can be considered as a filter design process(See Ref.\cite{22} for details).
Any kinds of control sequence can be expressed as a particular filter function. Analytical expression for CPMG sequence with pulses of even number N(replace $ \sin^2{x} $ with $ \cos^2{x} $ for odd-N CPMG) embedded before the measurement point can be written as:
\begin{equation}
F(x)=8\sin^4(\dfrac{x}{4N})\sin^2(\dfrac{x}{2})/\cos^2{\dfrac{z}{2N}}
\end{equation}
Note that the intervals between pulses are fixed in our scheme. Namely, N is proportional to the length of the fiber. As the particular property of CPMG described above, we can investigate the dynamical evaluation process using the deduced filter function:
\begin{equation}
F(\omega L)=8\sin^4(\dfrac{\omega}{4n})\sin^2(\dfrac{\omega L}{2})/\cos^2{\dfrac{\omega}{2n}}
\end{equation}

\begin{figure}
\includegraphics[width=8.5cm]{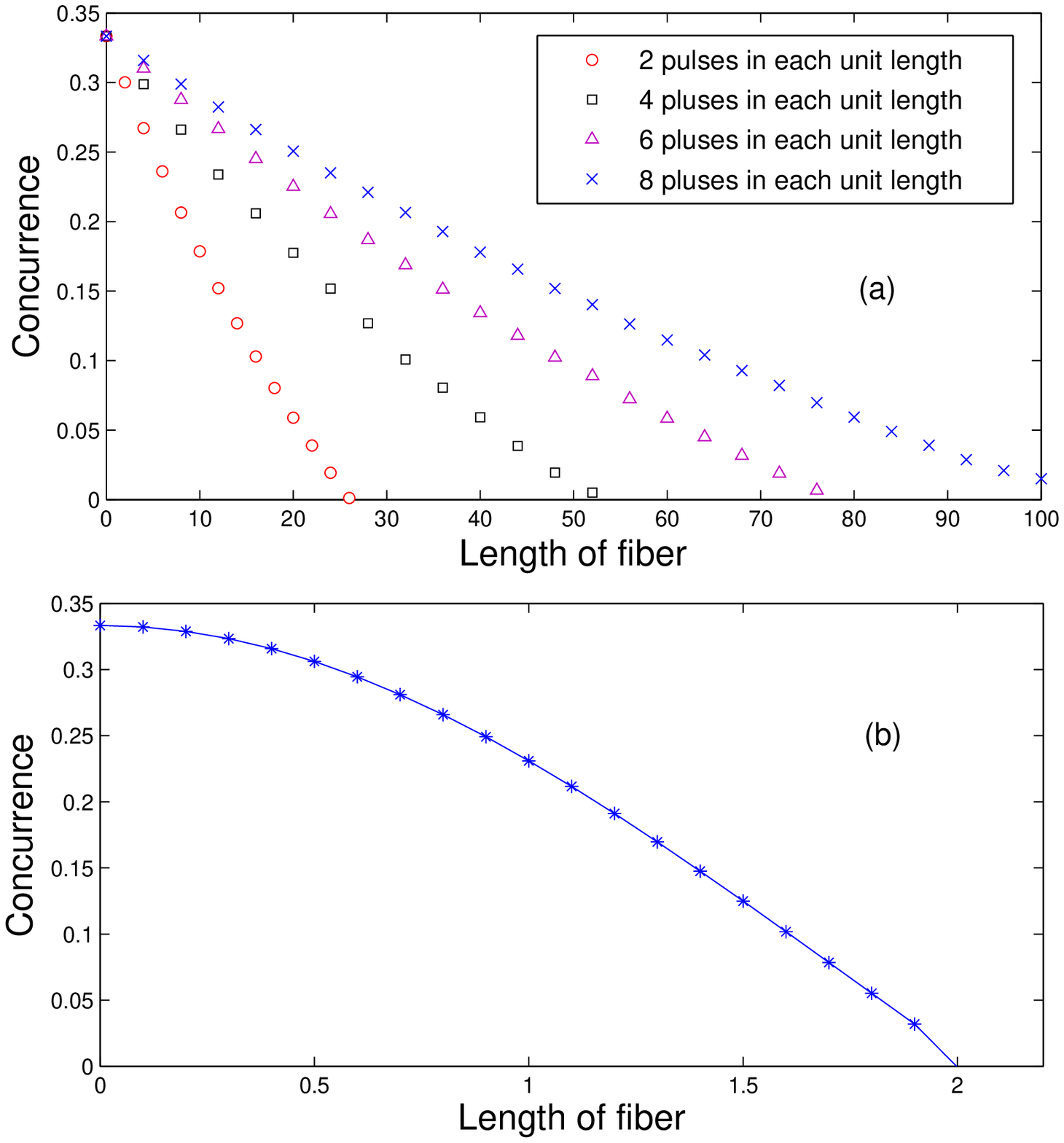}
\caption{(a)Evoluation of entanglement along the fiber for different
pulse intervals. (b)Evoluation of entanglement with one SE pulse
embedded into the fiber. Note that in SE case here, the point at which the pulse is
embedded changes with the measurement point.}
\end{figure}

\begin{figure}
\includegraphics[width=8.5cm]{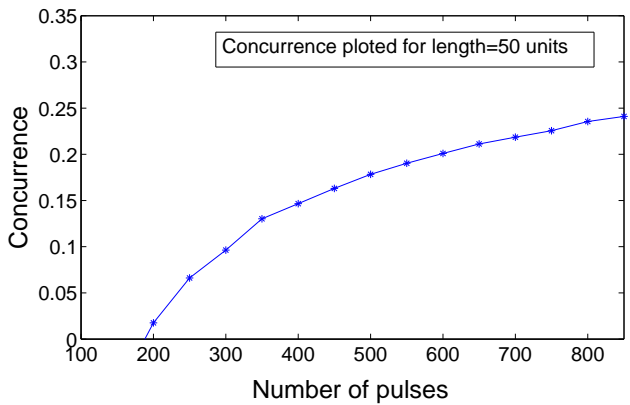}
\caption{Concurrence as a function of the total pulse number within
50 units length. }
\end{figure}

Consider the frequency-dependent character of the evaluation process. From equation (8) it can be generated that the elements of the resulting density matrix increase with $ \sigma $. Namely, decoherence is inhibited, instead of strengthen, by the optical frequency dispersion. This effect, though can be hardly observed due to the ordinary situation with $ \sigma<<\omega_0 $, may more generally inspire us that under a DD control process an additional degree of freedom with special structure can stabilize the spin coherence. In the optical case here, since a significant decoherence will emerge within a length scale of $ \sigma^2f(L)<<1 $, equation (8) can then be simplified as:
\begin{equation}
\rho_{13,14,23,24}=\rho_{13,14,23,24}(0)e^{-\omega_0^2f(L)}
\end{equation}
which is not sensitive to the optical frequency dispersion.

We now turn to the characterization of the degree of entanglement of the two-photons state. Wootters concurrence\cite{24} is particularly convenient for the two-qubit case here. Other reliable measure of entanglement will yield the same conclusion. The concurrence can be calculated explicitly from the denity matrix $ \rho $ discribed in equation (12)
\begin{equation}
C(\rho)=\max(0,\sqrt{\lambda_1}-\sqrt{\lambda_2}-\sqrt{\lambda_3}-\sqrt{\lambda_4})
\end{equation}
where the quantities $ \lambda_i $ are the eigenvalues in decreasing order of the matrix:
\begin{equation}
\rho^{'}=\rho(\sigma^A_y\otimes\sigma^B_y)\rho^*(\sigma^A_y\otimes\sigma^B_y)
\end{equation}

In the following we will analysis the evoluation of entanglement of a class of bipartite density matrices initially prepared with the common X-form\cite{25}:
$$
\rho^{AB}=\begin{pmatrix}
\rho_{11} & 0 & 0 & \rho_{14}\\0 & \rho_{22} & \rho_{23} & 0\\0 & \rho_{32} & \rho_{33} & 0\\\rho_{41} & 0 & 0 & \rho_{44}
\end{pmatrix}
$$
which occur in many contexts including pure Bell states as well as Werner mixed states.For the concurrence with such arbitrary initial state
there is no compact analytical expression. However, it has been readily showed\cite{26} that entanglement sudden death always occur under a classic Gauss noise, as well as the evolution with DD control implemented. The DD method inhibit losing of entanglement by reshaping the exponential factor
of the diagonal elements. We now give the numerical simulation to evaluate the performance of the CPMG in our scheme.

Without losing universality, we numerical analysis the concurrence evolution with a initial specific X-form entangled state:
$$
\rho^{AB}(0)=\dfrac{1}{3}\begin{pmatrix}
1/2 & 0 & 0 & 0\\0 & 1 & 1 & 0\\0 & 1 & 1 & 0\\0 & 0 & 0 & 1/2
\end{pmatrix}
$$

Thus the initial concurrence is $ C(0)=\dfrac{1}{3} $. In Fig.2 (a)
we give the numerical simulation of the concurrence evolution under
a Gauss 1/f noise (the noise spectral density is $ S(\omega)\propto
1/\omega $, origination of which are discussed in Ref. \cite{27}).
We observed that the concurrence of entanglement improves as the
intervals between pulses decrease. As a contrast we also depicted
the entanglement evolution process with SE pulse implemented in
Fig.2 (b). It can be inferred that the CPMG sequence drastically
out-perform the single SE pulse in inhibiting the entanglement
sudden death. The performance of the SE pulse has been experimental
tested as an effective method for coherence control\cite{28}.

\begin{figure}
\includegraphics[width=8cm,height=5.8cm]{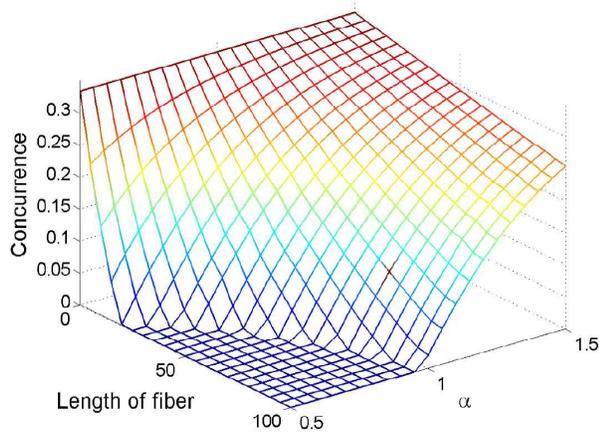}
\caption{Concurrence evolution under a general Gauss noise
environment with variable $ \alpha $ ranges from 0.5 to 1.5}
\end{figure}
We also analysis the entanglement preservation in a fixed fiber
length as the total number of pulses changes. Thus the minimum
number of pulses can be estimated to achieve a given level of
entanglement. Fig.3 shows that the entanglement rebirth while a certain
number of pulses are implemented. However,it can be seen that the
growth rate of concurrence declines while the number of pulses
increases. This makes a preservation of entanglement to a extremely
high level be difficult.

The performance of CPMG sequence under a general Gauss $ 1/f^\alpha
$ with spectral density $ S(\omega)\propto 1/\omega^\alpha $ is also
considered. Fig.4 shows the concurrence evolution under the noise
spectral density with variable $ \alpha $ ranges from 0.5 to 1.5.

In conclusion, we demonstrated that dynamic decoupling can effectively overcome the stochastic dispersion in the polarization-maintaining birefringent fibers under classic Gauss noises. Entanglement can be successful preserved by embedding waveplates as CPMG sequence into the fibers. Our work will evidently enhance the scope of fiber-based quantum communication. We also hope this method be extended with more complicated DD sequence to general decoherent environment in ordinary sigle mode fibers.

This work is supported by the National Basic Research Program (2011CB921200) and National Natural Science Foundation of China (Grants No. 60921091 and No. 10874162)


\begin{thebibliography}{unsrt}
\bibitem{1}C. Antonelli, M. Shtaif, and M. Brodsky, Phys. Rev. Lett. \textbf{106}, 080404 (2011).
\bibitem{2}J. H. Eberly and T. Yu, Science \textbf{316}, 555 (2007).
\bibitem{3}T. Yu and J. H. Eberly, Science \textbf{323}, 598 (2009).
\bibitem{4}P. Zanardi and M. Rasetti, Phys. Rev. Lett. \textbf{79}, 3306 (1997).
\bibitem{5}R. Prevedel, M. S. Tame, A. Stefanov, M. Paternostro, M. S. Kim, and A. Zeilinger, Phys. Rev. Lett. \textbf{99}, 250503 (2007).
\bibitem{6}P. W. Shor, Phys. Rev. A \textbf{52}, R2493 (1995).
\bibitem{7}N. Boulant, L. Viola, E. M. Fortunato, and D. G. Cory, Phys. Rev. Lett. \textbf{94}, 130501 (2005).
\bibitem{8}D. Vitali, P. Tombesi, and G. J. Milburn, Phys. Rev. Lett. \textbf{79}, 2442 (1997).
\bibitem{9}L. M. K. Vandersypen and I. L. Chuang, Rev. Mod. Phys. \textbf{76}, 1037 (2005).
\bibitem{10}J. F. Du, X. Rong, N. Zhao, Y. Wang, J. H. Yang and R.B. Liu, Nature \textbf{461}, 1265 (2009).
\bibitem{11}M. Mukhtar, W.T. Soh, T.B. Saw, and J.B. Gong, Phys. Rev. A \textbf{82}, 052338 (2010)
\bibitem{12}M. Mukhtar, T.B. Saw, W.T. Soh, and J.B. Gong, Phys. Rev. A \textbf{81}, 012331 (2010)
\bibitem{13}Y. Wang, X. Rong, P.B. Feng, W.J. Xu, B. Chong, J.H. Su, J.B. Gong, and J. F. Du, Phys. Rev. Lett. \textbf{106}, 040501 (2011)
\bibitem{14}S. Meiboom and D. Gill, Rev. Sci. Instrum. \textbf{29}, 668 (1958).
\bibitem{15}K. Khodjasteh and D. A. Lidar, Phys. Rev. Lett. \textbf{95}, 180501 (2005).
\bibitem{16}G. S. Uhrig, Phys. Rev. Lett. \textbf{98}, 100504 (2007).
\bibitem{17}J. R. West, B. H. Fong, and D. A. Lidar, Phys. Rev. Lett. \textbf{104}, 130501 (2010)
\bibitem{18}Lian-Ao Wu and Daniel A. Lidar, Phys. Rev. A \textbf{70}, 062310(2004)
\bibitem{19}Asoka Biswas1 and Daniel A. Lidar, Phys. Rev. A \textbf{74}, 062303(2006)
\bibitem{20}S. Damodarakurup, M. Lucamarini, G. Di Giuseppe, D. Vitali, and P. Tombesi, Phys. Rev. Lett. \textbf{103}, 040502 (2009).
\bibitem{21}M. Lucamarini, G. Di Giuseppe, S. Damodarakurup, D. Vitali, and P. Tombesi , Phys. Rev. A \textbf{83}, 032320 (2011).
\bibitem{22}L. Cywinski, Roman M. Lutchyn, Cody P. Nave, and S. Das Sarma, Phys. Rev. B \textbf{77}, 174509 (2008).
\bibitem{23}Phoenix S.Y. Poon and C.K. Law, Phys. Rev. A \textbf{77}, 032330 (2008).
\bibitem{24}W. K. Wootters, Phys. Rev. Lett. \textbf{80}, 2245 (1998).
\bibitem{25}T. Yu, J.H. Eberly, Quant. Inf. Comput. \textbf{7}, 459 (2007).
\bibitem{26}T. Yu, J.H. Eberly, Optics Communications \textbf{283}, 676 (2010).
\bibitem{27}J. Schriefl, Y. Makhlin, A. Shnirman, and G. Schon, New J. Phys. \textbf{8}, 1 (2006).
\bibitem{28}C. F. Li, J. S. Xu, X. Y. Xu, K. Li, and G. C. Guo, Nature Physics \textbf{7}, 752 (2011).

\end{thebibliography}
\end{document}